\documentclass{aajk}
\usepackage{graphicx}
\usepackage{txfonts}
\usepackage{natbib}
\bibpunct{(}{)}{;}{a}{}{,}   
\newcommand{\grsim}{\mathrel{\hbox{\rlap{\hbox{\lower4pt\hbox{$\sim$}}}\hbox{$>$}}}}

\newcommand{\tausync}{\tau_{\rm s}}

\newcommand{\eqb}{\begin{eqnarray}}
\newcommand{\eqe}{\end{eqnarray}}

\newcommand{\diff}{{\rm d}}
\newcommand{\sigmaT}{\sigma_{\rm T}}

\newcommand{\tauT}{\tau_{\rm T}}

\newcommand{\Tbright}{T_{\rm B}}

\newcommand{\nusynch}{\nu_{\rm s}}

\newcommand{\nuGHz}{\nu_{\rm GHz}}
\newcommand{\numax}{\nu_{\rm max}}
\newcommand{\rmtwo}{R_{-2}}
\newcommand{\doppler}{{\cal D}}
\newcommand{\tcool}{t_{\rm cool}}
\newcommand{\fluxic}{{\cal F}_{\rm IC}}
\newcommand{\fluxghz}{{\cal F}_{\rm GHz}}
\newcommand{\PKSfifteennineteen}{PKS~1519~$-$273} 
\newcommand{\PKSofourofive}{PKS~0405~$-$385} 
\newcommand{\Jeighteennineteen}{J~1819~$+$3845}
 
\begin{document}
\title{High brightness temperatures and circular polarisation in
  extra-galactic radio sources}
\author{J.G. Kirk
\and O. Tsang
}
\institute{Max-Planck-Institut f\"ur Kernphysik, Saupfercheckweg 1, 
D-69117 Heidelberg, Germany
}
\date{Received <date> / Accepted <date>}
\titlerunning{High brightness temperature radio sources}
\abstract
{Some rapidly variable extra-galactic radio sources show
very high brightness temperatures $T_{\rm B}>10^{12\,}$K
and high degrees 
of circular polarisation ($\sim1\%$). 
Standard synchrotron models that assume a power-law electron distribution 
cannot produce such high temperatures and 
have much
lower degrees of intrinsic circular polarisation.}
{We examine the synchrotron and inverse Compton radiation from a 
monoenergetic electron distribution and discuss
the constraints placed upon it by radio, optical and 
hard X-ray/gamma-ray observations.}
{The standard expressions of synchrotron theory are used. 
Observational constraints on the source parameters are
found by formulating the results 
as functions of the source size, Doppler boosting factor, 
optical depth to synchrotron self-absorption
and maximum frequency of synchrotron
emission, together with a parameter governing 
the strength of the inverse Compton radiation.}
{The model gives brightness temperatures 
$T_{\rm B}\sim10^{13}$ to
$10^{14}\,$K for
moderate ($\lesssim10$)
Doppler boosting factors 
together with
intrinsic 
degrees of circular 
polarisation at the percent level. 
It predicts 
a spectrum $I_\nu\propto \nu^{1/3}$ between the radio and
the infra-red as well as emission in the MeV to GeV range. 
If 
the energy density in relativistic
particles is comparable to or greater than the magnetic energy density,
we show that electrons do not cool within the source, enabling 
the GHz emission to emerge without absorption and 
the potentially catastrophic energy losses 
by inverse Compton scattering to be avoided.
Magnetically dominated sources can also fulfil these requirements 
at the cost of 
a slightly lower limit on the brightness temperature.}
{We suggest that sources such as
\PKSfifteennineteen, 
\PKSofourofive\ and \Jeighteennineteen\  
can be understood within this scenario without invoking 
high Doppler boosting factors, coherent emission mechanisms,
or the dominance of proton synchrotron radiation.}
\maketitle
\offprints{J.G. Kirk
}
\email{john.kirk@mpi-hd.mpg.de}
\date{Received \dots / Accepted \dots }
%

\section{Introduction}

The rapidly varying radio flux density observed 
in several extra-galactic sources implies 
a very high brightness temperature $\Tbright$ at the source
\citep{wagnerwitzel95}.
In cases such as \PKSfifteennineteen\ and \PKSofourofive,
where the variability is identified as interstellar scintillation 
\citep{macquartetal00,rickettetal02}
realistic models of the scattering screen 
require $T_{\rm B}>10^{13}\,$K. 
The recently discovered diffractive scintillation in 
\Jeighteennineteen\ 
requires $T_{\rm B}>2\times10^{14}\,$K \citep{macquartdebruyn05}. 
In other sources, the variability can be interpreted as  
intrinsic, in which case the implied temperature 
can be even higher.

Such high temperatures are difficult to understand within 
standard models
of synchrotron emission, which assume a power-law distribution of electrons. 
The 
energy radiated by inverse Compton scattering in the Thomson regime rises 
dramatically when the intrinsic brightness 
temperature exceeds a certain threshold, roughly 
equal to $3\times10^{11}\,$K
in the case of a power-law electron distribution, although 
a flaring source might exceed this limit for a short time
\citep{kellermannpaulinytoth69,slysh92,melrose02}. 
However,
as first pointed out by \citet{crusius-waetzel91}, 
the threshold temperature is higher for a monoenergetic 
electron distribution than for a power-law distribution, since, 
in the former case, 
photons of 
low frequency (compared to the characteristic synchrotron frequency)
can escape without absorption by low energy electrons. 

High brightness temperature sources frequently display circular polarisation
at the~$1\%$ level or above \citep{macquart03}.
Because this is much higher than the 
value $m_{\rm e}c^2/k_{\rm B}T_{\rm B}$ conventionally estimated   
for the intrinsic emission of a power-law electron distribution, 
propagation effects are the favoured explanation
\citep{jonesodell77a,wardleetal98,macquartmelrose00,beckertfalcke02,ruszkowskibegelman02,broderickblandford03,wardlehoman03}.
A monoenergetic electron model, on the other hand, 
predicts {\em intrinsic} circular polarisation at the $\%$ level, obviating
the need for a conversion process. Additional predictions are 
a hard $I_\nu\propto\nu^{1/3}$ synchrotron 
spectrum extending from the radio at least 
up to the infra red, and an inverse Compton component 
in the MeV to GeV range.

\section{Model parameters}
We consider an idealised model 
characterised by a single length scale $R$ and 
Doppler boosting factor $\doppler=\sqrt{1-\beta^2}/(1-\beta\cos\phi)$,
where $c\beta$ is the source speed with respect to the 
rest frame of the host galaxy and $\phi$ 
is the angle between the velocity and
the line of sight.
A homogeneous, isotropic, monoenergetic 
distribution of electrons with Lorentz factor $\gamma$
and number density $n$,
embedded in a uniform magnetic field $B$ is assumed 
in the rest frame of the source.
This captures the relevant properties of 
a power-law electron distribution with a lower energy \lq\lq cut-off\rq\rq, 
provided the synchrotron opacity at low frequency is dominated by electrons
whose energy is close to that of the cut-off, as is the case if the
distribution rises towards the cut-off sufficiently rapidly: 
$\diff \ln n/\diff\ln \gamma>-1/3$.
Quasi-monoenergetic distributions of this type 
can account for the lack of Faraday depolarisation in
parsec-scale emission regions \citep{wardle77,jonesodell77b} and 
have recently
been discussed in connection with statistical trends in the observed 
distribution of superluminal velocities as a function of observing frequency
and redshift \citep{gopal-krishnabiermannwiita04}. 
Monoenergetic models have been considered previously in connection 
with high brightness temperature sources by \citet{crusius-waetzel91},
and, more recently, by \citet{protheroe03},
who, however, adopted the assumption of equipartition between
particle and magnetic energy densities in the source and
introduced an additional, potentially large, 
geometrical factor.
\citet{slysh92} also treated a monoenergetic model, but did
not allow for  
the possibility of multiple Compton scatterings and further
restricted his treatment to optically thick sources.

Neglecting, for a moment, the angle between the line of sight and the
magnetic field, the source model contains five free parameters:
$R$, $\doppler$, $\gamma$, $B$ and $n$. Of these, only two 
are directly constrained by observations. 
Firstly, an upper limit on the linear dimension $R$, 
is given by the maximum permitted angular size
inferred from the presence of scintillation. Secondly,
surveys of superluminal motion \citep{cohenetal03} 
reveal apparent transverse velocities that 
are mostly less than $10c$, but extend up to 
$30c$.
These
data suggest that 
$\doppler\lesssim10$ for most sources, 
although it may be up to 30 or 40 in a few cases.
If the source variability arises intrinsically, rather than from 
scintillation, an upper limit to the quantity 
$R/\doppler^2$ is obtained.

Additional constraints can be
found by eliminating $\gamma$, $B$ and $n$ in favour of three
new parameters. 
The first of these is the optical depth to synchrotron self-absorption
$\tausync$.
Denoting the Thomson optical depth of
the monoenergetic electrons by $\tauT=n\sigmaT R$
(where $\sigmaT$ is the Thomson cross-section)
one has
\eqb
\tausync&=&{\sqrt{3}\,\tauT m_{\rm e} c^3 K_{5/3}(x)\over
8\pi e^2\nusynch\gamma^3}
\label{synchdepth}
\eqe
(e.g., \citep{Longair92}). Here, 
$K_{5/3}(x)$ is a modified Bessel function,
\eqb
x&=&\nu(1+z)/(\doppler\nusynch)
\label{xdef}
\eqe 
$\nu$ is the observing frequency,
$\nusynch=3eB\sin\theta\gamma^2/(4\pi m_{\rm e} c)$
is the characteristic synchrotron frequency,
for an angle 
$\theta$ between the line of sight and the magnetic field
$B$ (both measured in the frame in which the source is at rest)
and $z$ is the redshift of the host galaxy.
$\tausync$ 
is a convenient parameter because it controls the brightness temperature,
which has a single maximum close to $\tausync=1$, when the other 
parameters are fixed.

The second new parameter, $\xi$, determines the inverse Compton luminosity. 
Synchrotron  photons act as targets for inverse scattering by 
the same electron population that produced them, 
giving rise to a 
first generation of upscattered photons. These, in turn, act as targets 
off which the electrons  produce a second generation, and so forth.  
We define 
$\xi$ as the ratio of the energy densities (or,
equivalently, luminosities) in consecutive generations,
assuming the scattering processes 
occur in the Thomson regime 
\citep[e.g.,][]{melrose02}. 
This quantity is somewhat sensitive to 
the geometry and homogeneity of the source
\citep{protheroe02}, but, 
for a uniform, roughly spherical source that is optically thin to scattering,
one can write
\eqb
\xi&=&4\gamma^2\tauT/3
\label{xidef}
\eqe
since the photon frequency is increased on 
average by the factor $4\gamma^2/3$ per scattering, and 
the probability of scattering is $\tauT$.
The assumption of Thomson scattering leads to a divergent luminosity whenever
$\xi>1$, independent of the source of the initial (zeroth generation) photon
targets. This phenomenon has acquired the name {\em Compton Catastrophe}.
If the source is optically thin to most of the synchrotron photons, $\xi$ is
equal to the ratio of the energy density in these photons to that in the
magnetic field. This is, in fact, the most commonly used definition. 
However, using our
definition (\ref{xidef}) the condition 
for catastrophe remains $\xi>1$ in both the optically thin and thick 
(to synchrotron absorption) cases
as well as in the case where photons of the cosmic
microwave background provide a more effective target than do the 
synchrotron photons.
We show below that inverse Compton scattering indeed takes place at least
initially in the
Thomson regime. Therefore, in order to avoid 
inordinately large Compton losses, we require
\eqb
\xi&\lesssim 1
\label{iccond}
\eqe
which automatically ensures $\tauT\ll1$. 

The observed brightness temperature $T_{\rm B}$,
related to the specific intensity of radiation $I_\nu$ in the direction of
the source by $T_{\rm B}=c^2 I_\nu/(2\nu^2k_{\rm B})$, follows
straightforwardly from the solution of the 
equation of radiation transport:
\eqb
{k_{\rm B}T_{\rm B}\over m_{\rm e} c^2}
&=&
{\doppler\over 1+z}
\left({\gamma F(x)\over2 x^2 K_{5/3}(x)}\right)
\left(1-\textrm{e}^{-\tausync}\right)
\label{temp1}
\eqe
where $F(x)=x\int_x^\infty \diff t K_{5/3}(t)$ is the standard synchrotron
function in the Airy integral approximation.  
Eliminating the parameters $\tauT$, $\nusynch$ and $\gamma$   
using Eqs.~(\ref{synchdepth}), (\ref{xdef}) and
(\ref{xidef}), gives
\eqb
{k_{\rm B}T_{\rm B}\over m_{\rm e} c^2}
&=&
\left({3^{3/2}m_{\rm e} c^3\over 4^5\pi e^2\nu}\right)^{1/5}
\!\!
\left({\xi\doppler^6\over(1+z)^6}\right)^{1/5}
\!\!
\left({1-\textrm{e}^{-\tausync}\over \tausync^{1/5}}\right)
\!
\left({F(x)\over x^{9/5} K_{5/3}^{4/5}(x)}\right)
\label{temp2}
\eqe

The first term in parentheses on the right-hand side of this equation 
is independent of the source parameters. 
The second is constrained by observation to be $\lesssim10$. 
The third reaches a maximum
of the order of unity at $\tausync\sim1$. The fourth, however, diverges for 
small $x$ as $x^{-2/15}$. Thus, even with $\xi<1$ and $\doppler<10$, 
it is possible to 
choose an $x$ for which this formula gives 
an arbitrarily high brightness temperature at any specified 
observing frequency.  
However, this model predicts a synchrotron spectrum that rises with  frequency
at least 
as fast as $I_\nu\propto \nu^{1/3}$ between the observing frequency 
$\nu$ and the frequency $\nu/x$. In principle, this 
can be constrained by observation. 
For \PKSfifteennineteen\ and \PKSofourofive, for example, 
the modest optical fluxes
suggest that $\numax\lesssim10^{14}\,$Hz  
\citep[][and Wagner, priv.\ comm.]{heidtwagner96}, although 
it is not known whether these observations 
coincided with 
an episode of 
high brightness temperature radio emission in these variable
sources.
Nevertheless, for our third
parameter, we adopt the frequency $\numax=\nu/x$
above which the optically thin
synchrotron radiation cuts off.

\section{Brightness temperature and circular polarisation}
Rewriting the brightness temperature Eq.~(\ref{temp2}) in terms of 
the three new parameters $\tausync$, 
$\xi$ and $\numax$, expressed in convenient units, we find that
it is independent of the source size, and depends only weakly on $\xi$
and $\numax$:
\eqb
T_{\rm B}&=&1.2\times10^{14}
\left({\doppler_{10}^{6}\xi\over(1+z)^6}\right)^{1/5}
\left({1-\textrm{e}^{-\tausync}\over\tausync^{1/5}}\right)\nu_{{\rm max}14}^{2/15}
\nuGHz^{-1/3}\,
\textrm{K}
\label{bright1}
\eqe
where
$\nu_{{\rm max}14}=\numax/(10^{14}\,\textrm{Hz})$, 
$\nuGHz=\nu/(1\,\textrm{GHz})$, $\doppler_{10}=\doppler/10$, 
and the approximations 
$F(x)\approx 2.15 x^{1/3}$ and $K_{5/3}(x)\approx 1.43 x^{-5/3}$, 
valid for $x\ll 1$,  
are used.
According to this equation, brightness temperatures of up to 
roughly $10^{13}\,$K 
at GHz frequencies, as observed in 
\PKSfifteennineteen\ and \PKSofourofive, 
can be achieved with $\xi\lesssim1$ and $\doppler=1$. 
With 
$\doppler\approx 15$, 
the model permits $T_{\rm B}=2\times10^{14}\,$K, as observed in
\Jeighteennineteen.

In addition to the high brightness temperature,
a particularly interesting source property is the 
degree of intrinsic circular polarisation $r_{\rm c}$.
Assuming a pure electron-proton plasma that is optically thin to synchrotron 
self-absorption, this quantity is also independent of source size:
\eqb
r_{\rm c}&=&{1\over3\gamma}\left({2\over x}\right)^{1/3}\cot\theta\,
\Gamma(1/3)
\nonumber\\
&=&
0.019\times
\left({(1+z)\tausync\over\doppler_{10} \xi}\right)^{1/5}
\nu_{{\rm max}14}^{1/5}\,\cot\theta
\label{circular}
\eqe 
\citep{melrose80}.
In the case of a power-law electron distribution, $r_{\rm c}$ changes sign
when the optically thick regime is entered \citep{jonesodell77a}. We are not
aware of the
corresponding calculation for a monoenergetic distriubtion, 
but, to order of magnitude, one can estimate the 
peak value using this expression, which 
is remarkably insensitive to all source parameters
other than the magnetic field direction.
Several extra-galactic sources of extremely high brightness temperature 
display circular polarisation at the percent level
\citep{macquart03}, in particular 
PKS~1519~$-$273 and PKS~0405~$-$385. 
In the absence of a
low-energy
cut off in the electron distribution, $r_{\rm c}\sim1/\gamma$, which is 
far too small to explain the observations. However, Eq.~(\ref{circular})
shows that for a mono-energetic electron distribution, the 
intrinsic emission can be polarised at the $\%$ level, or above, 
depending on
the geometry of the magnetic field configuration. 

\section{Discussion}
The electron Lorentz factor implied by the above analysis 
is:
\eqb
\gamma&=&
2.8\times10^3
\left({\xi\doppler_{10}\nu_{{\rm max}14}^{2/3}\over
\tausync(1+z)}\right)^{1/5}
\nuGHz^{-1/3}
\label{gammavalue}
\eqe
A key ingredient of this model is the absence of electrons of lower 
Lorentz factor, since these 
would absorb the GHz emission, leading to a reduction
of the brightness temperature. Specifically, we require a quasi-monoenergetic
distribution such that $\diff \ln n/\diff\ln \gamma>-1/3$ at Lorentz factors
lower than that given by Eq.~(\ref{gammavalue}).
Such a distribution is not a natural consequence
of, for example, the first-order Fermi process
at relativistic shocks \citep[e.g.,][]{kirk05}. On the other hand, 
a relativistic 
thermal distribution, which rises at low energy as $\gamma^2$ 
is well-approximated by a monoenergetic distribution of energy roughly equal
to the temperature. The addition of a power-law tail to higher energy
would not change this conclusion.

The Lorentz factor implied by Eq.~(\ref{gammavalue})
is higher than the 
cut-off conventionally assumed
when modelling radio sources \citep[e.g.,][]{gopal-krishnabiermannwiita04}.
Nevertheless, scenarios exist
which suggest such values.
One example is 
an electron-proton jet with a bulk Lorentz
factor $\Gamma\sim 10$ which is
thermalised at a shock front. If the 
downstream electron and ion temperatures
are equal, the distribution 
can be approximated as monoenergetic with an electron
Lorentz factor of $\Gamma m_{\rm p}/m_{\rm e}$, where 
$m_{\rm p}$ and $m_{\rm e}$ are the proton and
electron masses. Another possibility
is that the electrons are accelerated at a current sheet 
in an electron-proton plasma in which the magnetic energy density
is comparable to the rest-mass energy density \citep{kirk04}.
Each of these possibilities relies on the composition
of the source plasma being electron-proton. Interestingly, so does 
the relatively high 
degree of intrinsic circular polarisation given by Eq.~(\ref{circular}).  

Although it is conceivable that continuous re-acceleration 
prevents the accumulation of low energy electrons, both 
the current sheet and the shock scenario 
envisage a finite escape rate of particles from the acceleration
or thermalisation region. 
Escaping particles subsequently 
cool by synchrotron and inverse Compton emission. Therefore, the 
model electron distribution is self-consistent only if
these particles 
can be evacuated from the source in a time shorter than the cooling
timescale. 
The ratio of the 
electron cooling time $\tcool$ 
to the light-crossing time of the source 
can be written as:
\eqb
{c \tcool/ R}&=&{\eta/\xi}
\label{crossing}
\eqe
where $\eta$ is the ratio of the energy density in relativistic electrons
to that in the magnetic field. 
Writing $R=0.01\rmtwo\,$parsec, we find: 
\eqb
\eta&=&
{\gamma n m c^2\over \left(B^2/8\pi\right)}
\,=\,                         
2.9\left({\doppler_{10}^{13}
\xi^8\over(1+z)^{13}\tausync^3\nu_{{\rm max}14}^{8}}\right)^{1/5}
\nuGHz^{-1}\rmtwo^{-1}\sin^2\theta
\label{etadef}
\eqe
Clearly, very small sources tend to be particle dominated and, since $\xi<1$, 
they satisfy the self-consistency requirement $c\tcool/R>1$. However, 
Eq.~(\ref{etadef}) 
shows that $\eta$ is also quite 
sensitive to $\doppler$, $\xi$ and $\numax$, so that
magnetically dominated sources are by no means ruled out, provided they
have $\xi\lesssim\eta$. Since the brightness temperature is proportional to
$\xi^{1/5}$ (Eq.~\ref{bright1}) it is slightly lower for
magnetically dominated sources.
 
Although lacking a compelling physical justification, 
the assumption of equipartition, $\eta=1$, can be used to define
an 
\lq\lq equipartition Doppler factor\rq\rq. This leads,
in the standard model, to a relatively low limit on the 
brightness temperature 
$T_{\rm B, eq}\lesssim 3\times10^{10}\,$K
\citep{singalgopalkrishna85,readhead94}, which 
has some observational support \citep{cohenetal03}. 
In the monoenergetic model, however, this 
assumption
re-introduces a dependence on the source size, but does not 
substantially constrain the brightness temperature, as can be seen
from equations (\ref{bright1}) and (\ref{etadef}).

The parameter
\eqb
\zeta&=&{\gamma h\nusynch\over m_{\rm e}c^2}
\,=\,
2.3\times10^{-4}\left({(1+z)^4\xi\over\tausync\doppler_{10}^{4}}\right)^{1/5}
\nu_{{\rm max}14}^{17/15}
\nuGHz^{-1/3}
\eqe
which gives the ratio of the energy of a photon of the 
characteristic synchrotron frequency to the electron rest-mass, as seen in the
rest frame of a relativistic electron is also independent of source size.
For $\zeta\ll1$, the first inverse Compton scattering 
takes place in the Thomson regime. In this case, it is
consistent to require $\xi\lesssim1$ 
in order to avoid an excessively large energy demand on the source
i.e, in order to avoid the Compton Catastrophe.
The first generation of inverse Compton photons has a frequency 
of approximately
\eqb
\nu_1&\approx&
4.3\left({\doppler_{10}^{2}\xi^2\over(1+z)^2\tausync^2}\right)^{1/5}
\nu_{{\rm max}14}^{19/15}
\nuGHz^{-2/3}\,\textrm{MeV}
\eqe
and its flux can be estimated to be
\eqb
\fluxic
&\approx&4.5\times10^{-6}
\left({(1+z)^3\xi^3\tausync^2\over\doppler_{10}^{2}}\right)^{1/5}
\nu_{{\rm max}14}^{1/15}
\nuGHz^{1/3}\fluxghz
\eqe
where $\fluxghz$ is the flux observed in the radio
at frequency $\nuGHz\,$GHz.
This estimate of the 
inverse Compton flux is generally 
above the detection threshold of instruments on 
the INTEGRAL satellite, as noted by \citet{protheroe03}.
Subsequent generations of inverse Compton scattered photons
are likely to fall into the Klein-Nishina regime, and
the maximum photon energy achieved by multiple inverse Compton scattering
is ultimately
limited by the electron energy, as seen in the observer's frame, which takes
the value:
\eqb
{\doppler\gamma mc^2\over 1+z}&=&
14
\left({\doppler_{10}^6\xi\over(1+z)^6\tausync}\right)^{1/5}
\nu_{{\rm max}14}^{2/15}
\nuGHz^{-1/3}\,\textrm{GeV}
\eqe

To summarise, synchrotron radiation from a 
monoenergetic electron distribution reproduces the extremely high 
brightness temperatures observed in variable extra-galactic radio sources,
and explains the observed levels of circular polarisation.
Therefore, it does not appear necessary to appeal to 
coherent mechanisms
\citep{krishnanwiita90,benfordlesch98,begelmanetal05} 
or to proton synchrotron radiation 
\citep{kardashev00} to understand these objects. 
Testable predictions of the theory are 
a hard radio to infra-red spectrum and gamma-ray emission in the
MeV to GeV range.
%
%

\begin{thebibliography}{33}
\expandafter\ifx\csname natexlab\endcsname\relax\def\natexlab#1{#1}\fi

\bibitem[{{Beckert} \& {Falcke}(2002)}]{beckertfalcke02}
{Beckert}, T. \& {Falcke}, H. 2002, \aap, 388, 1106

\bibitem[{{Begelman} {et~al.}(2005){Begelman}, {Ergun}, \&
  {Rees}}]{begelmanetal05}
{Begelman}, M.~C., {Ergun}, R.~E., \& {Rees}, M.~J. 2005, \apj, 625, 51

\bibitem[{{Benford} \& {Lesch}(1998)}]{benfordlesch98}
{Benford}, G. \& {Lesch}, H. 1998, \mnras, 301, 414

\bibitem[{{Broderick} \& {Blandford}(2003)}]{broderickblandford03}
{Broderick}, A.~E. \& {Blandford}, R.~D. 2003, American Astronomical Society
  Meeting Abstracts, 203,

\bibitem[{{Cohen} {et~al.}(2003){Cohen}, {Russo}, {Homan}, {Kellermann},
  {Lister}, {Vermeulen}, {Ros}, \& {Zensus}}]{cohenetal03}
{Cohen}, M.~H., {Russo}, M.~A., {Homan}, D.~C., {et~al.} 2003, in ASP Conf.
  Ser. 300: Radio Astronomy at the Fringe, 177

\bibitem[{{Crusius-Waetzel}(1991)}]{crusius-waetzel91}
{Crusius-Waetzel}, A.~R. 1991, \aap, 251, L5

\bibitem[{{Gopal-Krishna} {et~al.}(2004){Gopal-Krishna}, {Biermann}, \&
  {Wiita}}]{gopal-krishnabiermannwiita04}
{Gopal-Krishna}, {Biermann}, P.~L., \& {Wiita}, P.~J. 2004, \apjl, 603, L9

\bibitem[{{Heidt} \& {Wagner}(1996)}]{heidtwagner96}
{Heidt}, J. \& {Wagner}, S.~J. 1996, \aap, 305, 42

\bibitem[{{Jones} \& {O'Dell}(1977{\natexlab{a}})}]{jonesodell77b}
{Jones}, T.~W. \& {O'Dell}, S.~L. 1977{\natexlab{a}}, \aap, 61, 291

\bibitem[{{Jones} \& {O'Dell}(1977{\natexlab{b}})}]{jonesodell77a}
{Jones}, T.~W. \& {O'Dell}, S.~L. 1977{\natexlab{b}}, \apj, 214, 522

\bibitem[{{Kardashev}(2000)}]{kardashev00}
{Kardashev}, N.~S. 2000, Astronomy Reports, 44, 719

\bibitem[{{Kellermann} \& {Pauliny-Toth}(1969)}]{kellermannpaulinytoth69}
{Kellermann}, K.~I. \& {Pauliny-Toth}, I.~I.~K. 1969, \apjl, 155, L71

\bibitem[{{Kirk}(2004)}]{kirk04}
{Kirk}, J.~G. 2004, Physical Review Letters, 92, 181101

\bibitem[{{Kirk}(2005)}]{kirk05}
{Kirk}, J.~G. 2005, in X-Ray and Radio Connections (eds. L.O. Sjouwerman and
  K.K Dyer) Published electronically by NRAO,
  http://www.aoc.nrao.edu/events/xraydio Held 3-6 February 2004 in Santa Fe,
  New Mexico, USA, (E1.01) 6 pages

\bibitem[{{Krishnan} \& {Wiita}(1990)}]{krishnanwiita90}
{Krishnan}, V. \& {Wiita}, P.~J. 1990, \mnras, 246, 597

\bibitem[{{Longair}(1992)}]{Longair92}
{Longair}, M.~S. 1992, {High energy astrophysics. Vol.1: Particles, photons and
  their detection} (Cambridge: Cambridge University Press, 1992, 2nd ed.)

\bibitem[{{Macquart}(2003)}]{macquart03}
{Macquart}, J.-P. 2003, New Astronomy Review, 47, 609

\bibitem[{{Macquart} \& {de Bruyn}(2005)}]{macquartdebruyn05}
{Macquart}, J.-P. \& {de Bruyn}, G. 2005, A\&A in press, astro-ph/0510495

\bibitem[{{Macquart} {et~al.}(2000){Macquart}, {Kedziora-Chudczer}, {Rayner},
  \& {Jauncey}}]{macquartetal00}
{Macquart}, J.-P., {Kedziora-Chudczer}, L., {Rayner}, D.~P., \& {Jauncey},
  D.~L. 2000, \apj, 538, 623

\bibitem[{{Macquart} \& {Melrose}(2000)}]{macquartmelrose00}
{Macquart}, J.-P. \& {Melrose}, D.~B. 2000, \apj, 545, 798

\bibitem[{{Melrose}(1980)}]{melrose80}
{Melrose}, D.~B. 1980, {Plasma astrohysics. Nonthermal processes in diffuse
  magnetized plasmas - Vol.1: The emission, absorption and transfer of waves in
  plasmas} (New York: Gordon and Breach, 1980)

\bibitem[{{Melrose}(2002)}]{melrose02}
{Melrose}, D.~B. 2002, Publications of the Astronomical Society of Australia,
  19, 34

\bibitem[{{Protheroe}(2002)}]{protheroe02}
{Protheroe}, R.~J. 2002, Publications of the Astronomical Society of Australia,
  19, 486

\bibitem[{{Protheroe}(2003)}]{protheroe03}
{Protheroe}, R.~J. 2003, \mnras, 341, 230

\bibitem[{{Readhead}(1994)}]{readhead94}
{Readhead}, A.~C.~S. 1994, \apj, 426, 51

\bibitem[{{Rickett} {et~al.}(2002){Rickett}, {Kedziora-Chudczer}, \&
  {Jauncey}}]{rickettetal02}
{Rickett}, B.~J., {Kedziora-Chudczer}, L., \& {Jauncey}, D.~L. 2002, \apj, 581,
  103

\bibitem[{{Ruszkowski} \& {Begelman}(2002)}]{ruszkowskibegelman02}
{Ruszkowski}, M. \& {Begelman}, M.~C. 2002, \apj, 573, 485

\bibitem[{{Singal} \& {Gopal-Krishna}(1985)}]{singalgopalkrishna85}
{Singal}, K.~A. \& {Gopal-Krishna}. 1985, \mnras, 215, 383

\bibitem[{{Slysh}(1992)}]{slysh92}
{Slysh}, V.~I. 1992, \apj, 391, 453

\bibitem[{{Wagner} \& {Witzel}(1995)}]{wagnerwitzel95}
{Wagner}, S.~J. \& {Witzel}, A. 1995, \araa, 33, 163

\bibitem[{{Wardle}(1977)}]{wardle77}
{Wardle}, J.~F.~C. 1977, \nat, 269, 563

\bibitem[{{Wardle} \& {Homan}(2003)}]{wardlehoman03}
{Wardle}, J.~F.~C. \& {Homan}, D.~C. 2003, \apss, 288, 143

\bibitem[{{Wardle} {et~al.}(1998){Wardle}, {Homan}, {Ojha}, \&
  {Roberts}}]{wardleetal98}
{Wardle}, J.~F.~C., {Homan}, D.~C., {Ojha}, R., \& {Roberts}, D.~H. 1998, \nat,
  395, 457

\end{thebibliography}
%

%
\end{document}